\documentclass[aps,a4paper,10pt,twocolumn,nofootinbib]{revtex4}
\usepackage{amsmath}
\usepackage{amsfonts}
\usepackage{amssymb}
\usepackage{mathrsfs}
\usepackage[mathscr]{euscript}
\usepackage[dvips]{graphicx}
\usepackage{fancyhdr}
\usepackage{makeidx}
\usepackage{colordvi}

\def\overstrike#1#2{{\setbox0\hbox{$#2$}\hbox to \wd0{\hss
    $#1$\hss}\kern-\wd0\box0}}

\newcommand{\XDOI}[1]{\href{http://dx.doi.org/#1}{doi:#1}}

\begin{document}

\title{Comment on: Power loss and electromagnetic energy density in a dispersive metamaterial medium}

\author{P. Kinsler}
\email{Dr.Paul.Kinsler@physics.org}

\affiliation{
  Blackett Laboratory, Imperial College London,
  Prince Consort Road,
  London SW7 2AZ,
  United Kingdom.}

\begin{abstract}

By clarifying the approach of Luan \cite{Luan-2009pre}, 
 we can generalize the analysis of dispersive (meta)materials, 
 and treat other material responses involving not only loss, 
 but also gain and coherent response.

\end{abstract}

\lhead{\includegraphics[height=5mm,angle=0]{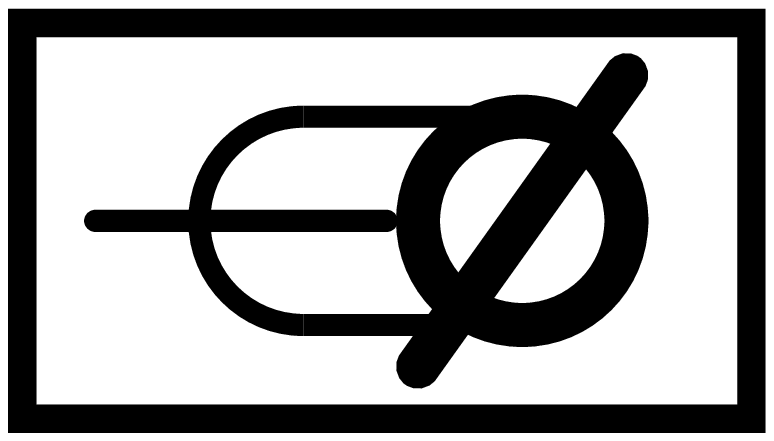}~~LUANCOM}
\rhead{
\href{mailto:Dr.Paul.Kinsler@physics.org}{Dr.Paul.Kinsler@physics.org}\\
\href{http://www.kinsler.org/physics/}{http://www.kinsler.org/physics/}
}

\date{\today}
\maketitle
\thispagestyle{fancy}

In \cite{Luan-2009pre}, 
 Luan treats the loss and energy density
 in dispersive media with potentially
 negative-valued permittivity $\epsilon$ and permeability $\mu$.
In the eqn. (L10) 
 (i.e. Luan's eqn. (10)), 
 the Poynting theorem is given as
~
\begin{align}
 -
  \nabla \cdot \left( \Vec{E} \times \Vec{H} \right)
&=
  \partial_t
  \left[
    \frac{\epsilon_0}{2}
    \Vec{E}^2 
   +
    \frac{\mu_0}{2}
    \Vec{H}^2 
  \right]
 +
  \Vec{E} \cdot \partial_t \Vec{P}
 +
  \Vec{H} \cdot \partial_t \Vec{M}
,
\label{eqn-Poynting}
\end{align}
 although I use $\partial_t \equiv d/dt$ rather than the overdot.
Here the electric field energy is $W_E= \epsilon_0 \Vec{E}^2$
 and the magnetic induction field energy is $W_H = \mu_0 \Vec{H}^2$.
The temporal increase or decrease in these 
 is balanced by either energy flow $\nabla \cdot (\Vec{E} \times \Vec{H})$
 or the polarization and magnetization ``residual''\cite{Kinsler-FM-2009ejp}
 terms 
 $R_e = \Vec{E} \cdot \partial_t \Vec{P}$
 and $R_h = \Vec{H} \cdot \partial_t \Vec{M}$.
Next,
 Luan considers \emph{losses} 
 in order to understand how $R_e, R_h$ can be divided into 
 physically meaningful components.

\emph{In this Comment} I point out that loss
 does not need to be considered so early in the argument -- 
 simply following the rest of Luan's procedure allows 
 the physical meanings of each component to arise naturally.

The residual terms  $R_e, R_h$ 
 neither have the form of a divergence,
 nor of an energy density change -- 
 instead
 they represent energy exchange (in either direction)
 between the field and the medium; 
 but since the Poynting theorem does not include an explicit 
 representation of energy stored in the medium
 (whether microscopic or not), 
 no term for the energy density of the medium is present.
Of course an energy exchange (``leakage'') 
 out of the field could be loss,
 but it might also be a temporarily storage.

Luan uses a plasmonic model for the polarization 
 (as per eqns. (L3) and (L7)),
 so that the electric residual $R_e$ from eqn. (L12)
 can be separated into two sub-components, 
~
\begin{align}
  R_e
&=
  \partial_t 
  \left[ 
    \frac{\Vec{P}^2}
         {2\omega_p^2\epsilon_0}
  \right]
 +
  \frac{\nu}{\omega_p^2 \epsilon_0}
   \left( \partial_t \Vec{P} \right)^2
 =
  \partial_t \left[ W_p \right]
 ~
 +
  R_p
.
\label{eqn-Re}
\end{align}
The magnetization follows a non-Lorentzian ``F-model'' response
 (as per eqns. (L4) and (L8)).
This model is in fact not strictly causal, 
 as discussed in \cite{Kinsler-2011ejp}, 
 although in practise the discrepancy can usually be neglected.
In any case, 
 its residual $R_h$ is given by eqn. (L14)
 and what follows in (L15-17).
It can also be divided up --
~
\begin{align}
  R_h
&=
  \mu_0
  \partial_t 
  \left[
   - 
    \frac{F}{2}
    \Vec{H}^2
   +
    \frac{1}{2\omega_0^2 F}
    \left(
      \partial_t \Vec{M} + F \partial_t \Vec{H} + \gamma \Vec{M}
    \right)^2
  \right]
\nonumber
\\
&\qquad
 +
  \frac{\gamma \mu_0 \Vec{M}^2}
       {F}
 =
  \partial_t \left[ W_m \right]
 ~~
 +
  R_m
.
\label{eqn-Rh}
\end{align}


\emph{First,} 
 there are terms that are true time derivatives, 
 i.e. they both look like and act like energy density terms
 in eqn. \eqref{eqn-Poynting}, 
 and so we can might call them the 
 polarization energy density ($W_p$) and
 magnetization energy density ($W_m$).
Luan's ``electric energy density'' is just the sum 
 of the electric field and polarization energy densities, 
 i.e. $W_e = W_E + W_p$; 
 similarly his ``magnetic energy density'' is the sum 
 of the magnetic field and magnetization energy densities, 
 i.e. $W_b = W_H + W_m$.
The total electromagnetic energy density is
 $W_{total}=W_e+W_h+W_p+W_m$; 
 and as would normally be expected, 
 neither the total nor any of its components 
 are negative.

\emph{Second,} 
 there are the remaining terms $R_p$ and $R_m$ which 
 neither have the form of a divergence or of an energy density change; 
 they therefore still represent energy exchange
 between the electromagnetic system and 
 some non-explicit part of the medium response.
Now, 
 however, 
 the energy density of the electromagnetic system 
 is $W_{total}$ -- 
 it is not just field energy density $W_{f}=W_E+W_H$.

\emph{Finally,}
 having divided up the residual terms $R_e$ and $R_h$, 
 we can now consider what the remaining energy exchange 
 terms $R_p$ and $R_m$ represent.
And,  
 in Luan's example, 
 they are losses and match perfectly with eqn. (L11) --
 but there is no requirement to attempt such a match in advance.
E.g. 
 we can introduce gain into Luan's electric response by 
 changing the sign of $\nu$, 
 and gain into the magnetic response 
 by changing the sign of $F$.
Indeed, 
 we might follow Luan's strategy with other types of material response, 
 and so get alternative residual terms:
 some parts of these might indeed be loss (as here), 
 but they might equally well be something else, 
 and represent (e.g.) coherent energy exchange with the medium.

%


\noindent
\emph{Acknowledgement:}
I acknowledge financial support from the EPRSRC (EP/E031463/1).

%


\end{document}